\documentstyle[sprocl,epsf]{article}

\def\Hth#1#2#3#4#5#6#7{}

\newcommand{\be}[1]{\begin{equation}\label{#1}}
\newcommand{\ee}{\end{equation}}

\newcommand{\prlput}[1]{}
\newcommand{\rem}[1]{}
\newcommand{\FFig}[1]{Fig.~{\ref{fig:#1}}}
\newcommand{\eref}[1]{(\ref{#1})}
\newcommand{\Eref}[1]{Eq.~(\ref{#1})}

\newcommand{\SSEC}[1]{Sec.~\ref{ssec:#1}}

\newcommand{\NN}{{\cal N}}
\newcommand{\DD}{{\cal D}}

\def\Lgr{{\cal L}}
\def\rarr{\rightarrow}
\def\eps{\epsilon}
\def\ta{\theta}
\def\Qt{\tilde Q}
\def\mn{{\mu\nu}}

\def\susy{supersymmetry}
\newcommand{\vev}[1]{\langle#1\rangle}

\def\half{{1\over 2}}
\def\NN{{\cal N}}

\def\none{$\NN=1$}
\def\ntwo{$\NN=2$}

\def\nfour{$\NN=4$}
\def\susy{supersymmetry}

\def\rarr{\rightarrow}

\begin{document}


\rightline{IASSNS-HEP-99/115}

\title{CONFINING PHASE OF THREE DIMENSIONAL 
SUPERSYMMETRIC QUANTUM ELECTRODYNAMICS}

\author{MATTHEW 
J.~STRASSLER}

\address{School of Natural Sciences\\Institute for Advanced Study \\ 
Olden Lane\\
Princeton, NJ 08540, USA\\E-mail: strasslr@ias.edu}


\maketitle\abstracts{
Abelian theories in three dimensions can have linearly confining
phases as a result of monopole-instantons, as shown, for $SU(2)$
Yang-Mills theory broken to its abelian subgroup, by Polyakov.  In
this article the generalization of this phase for \ntwo\
supersymmetric abelian theories is identified, using a dual
description.  Topologically stable BPS-saturated and unsaturated
particle and string solitons play essential roles.  A plasma of chiral
monopoles of charge 1 {\it and} -1 (along with their antichiral
conjugates) are required for a stable confining vacuum. \ntwo\ $SU(2)$
Yang-Mills theory broken to $U(1)$ lacks this phase because its chiral
monopoles all have the same charge, leading to a runaway instability.
The possibility of analogous confining phases of string theory, and a
dual field theoretic model thereof, are briefly discussed.
}





\section{Introduction}

Yuri Golfand was one of the first to construct supersymmetric versions
of abelian gauge theories.\cite{golfand,gersak}  The three dimensional
version of supersymmetric quantum electrodynamics (SQED) turns out to
be a very rich dynamical system whose non-perturbative properties have
recently received much
attention~\cite{nsewddd,kinsddd,ntwovort,ntwobrane,FT} and are still
being explored.  It seems fitting that the presentation of the
linearly confining regime of this theory, generalizing the work of
Polyakov in the non-supersymmetric context,\cite{polymono} be part of
this volume.

Duality has long been a powerful tool in studying confining gauge
theories.  't Hooft, Mandelstam and others discovered weakly coupled
field theories with solitonic flux tubes that behave somewhat
similarly to the tubes of chromoelectric flux which occur in QCD.  In
three dimensions, where charged particles are naively confined by a
logarithmically growing potential, it is still interesting to find
phases with the potential grows linearly.  In pure $SU(2)$ Yang-Mills
theory broken to its abelian subgroup, Polyakov showed, using the
duality of gauge fields and compact scalar fields in three dimensions,
that 't Hooft-Polyakov monopoles acting as dynamical instantons in
three dimensions lead to linear confinement.\cite{polymono}

The supersymmetric generalization of Polyakov's result does not give a
confining phase.  As Affleck, Harvey and Witten showed,\cite{AHW} the
monopole instantons in \none\ and \ntwo\ supersymmetric $SU(2)$
Yang-Mills theories cause the theory to develop an unstable potential.
The resulting instability drives the scale of the $SU(2)$-to-$U(1)$
breaking off to infinity, where the theory is Gaussian. No linearly
confining phase has been found for these theories.

Three dimensional gauge theories also have conformal fixed points,
where naive logarithmic confinement is lost and charged sources have a
$1/r$ potential.  With a large number of charged matter fields, it can
be shown using large $N_f$ techniques that many gauge theories have
this property.  Such fixed points are found in many supersymmetric
examples even when the number of charged matter fields is small.  For
\nfour\ supersymmetry, all abelian gauge theories are believed to have
conformal fixed points at the origin of moduli space.\cite{kinsddd}
This extends to a very wide class of \ntwo\ theories as well,
including the theory central to this paper, which contains a photon,
an electron, a positron, and their \ntwo\ superpartners.  The
discovery of a new dual description at these fixed
points,\cite{kinsddd} one which is distinct from the gauge field/scalar
duality used by Polyakov, permits new insights into these gauge
theories.

In this letter it will be shown that \ntwo\ SQED has a confining
phase, in analogy to other abelian non-supersymmetric examples.  The
mechanism is again that of Polyakov.  It will be explained below why
this mechanism does not work in the \ntwo\ $SU(2)$ Yang-Mills
theories: a stable confining phase in a supersymmetric theory requires
monopoles with charges 1 and -1, along with their antimonopoles, but
in $SU(2)$ one obtains only monopoles with charge 1 and antimonopoles
of charge -1, and a stable phase is not obtained.  It is also noted
that string theory may well have phases in which two-form flux does
not propagate freely and strings are linearly confined by domain walls
of two-form flux.   String duals of such phases
would be interesting to explore.

\section{Preliminaries: The Pure $U(1)$ Gauge Theory}

Let us begin with a trivial analysis of the theory without matter.
Consider \ntwo\ supersymmetric $U(1)$ gauge theory,\footnote{The
superfield language used here is that of \none\ \susy\ in four
dimensions, of which three-dimensional \ntwo\ is the dimensional
reduction.} consisting of a single $U(1)$ vector multiplet $V$, which
contains a photon, a photino, and a scalar $\phi$.  The field strength
$F^{\mu\nu}$ is contained in the gauge invariant multiplet
$\Sigma=\eps_{\alpha\beta}D^\alpha \bar D^\beta V.$ The Lagrangian of
the theory is $ \Lgr = \int d^4\ta\ {1\over 4g^2}\Sigma^2.  $

\subsection{The logarithmically confining phase}

Classically the theory has a moduli space of vacua given by
$\vev{\phi}$.  Quantum mechanically things are more interesting,
because we may replace the gauge field $A^\mu$ by its electromagnetic
dual, a scalar $\tau$ periodic under $\tau\rarr\tau+2\pi$.  We may
define a chiral superfield $T$ whose lowest component is $\phi/g^2+
i\tau$; then the full moduli space is the cylinder defined by
$\vev{T}$, with a trivial metric, or equivalently the plane defined by
$\vev{e^T}$.

More explicitly, given any effective action $S_{eff}$ for $\Sigma$, we
may introduce $T$ as follows. The path integral over $V$ can be
replaced with a path integral over $\Sigma$ only if a Lagrange
multiplier, a chiral superfield $T$, is added to implement the Bianchi
identity via the terms~\cite{lindrocek} $\Delta\Lgr = \int d^2\theta
T\bar D^2\Sigma + c.c.$.  Integrating by parts, we have
\be{partntwo}
\int\ \DD \Sigma\ \DD T 
\exp\left(iS_{eff}(\Sigma)+{i\over 4\pi}\int d^3x d^4\theta\ \Sigma 
(T+T^\dagger)\right)\ .
\ee
By integrating out $\Sigma$, we obtain an effective action for $T$.
For the free $U(1)$ theory, this effective action is trivial:
$\Lgr=\int\ d^4\ta\ (g^2/8\pi^2) T^\dagger T$.  Generally, however,
the field $T$, although gauge invariant, is non-local.  However, $e^T$
may be a local gauge invariant operator.\footnote{This has not
actually been proven, although mirror symmetry~\cite{kinsddd,ntwovort}
implies it must be true.\cite{FT}} If one inserts $e^T(x)$ into the
above path integral, the Bianchi identity is violated by a delta
function, $dF = 2\pi\delta(x)$; thus $e^T$ represents a pointlike instanton in
the form of a Dirac magnetic monopole.  (The normalization is chosen so
that the Dirac quantization condition is satisfied.)  We therefore
identify $e^T$ and $e^{-T}$ with chiral operators $M,\tilde M$ of
monopole charge $1,-1$, and their complex conjugates with
the conjugate antichiral operators $M^\dagger,\tilde M^\dagger$ of charge
$-1,1$.

Since the field $T$ takes expectation values on a cylinder, the
fundamental group $\pi_1$ of the moduli space is non-zero and there
exists the possibility of particle-like vortex solitons, or more
precisely, of solutions to the equations of motion which have winding
number.  The configuration $\vev{T(r,\theta,t)}\propto iq\ta$, $q$ an
integer, realizes this possibility.  Unfortunately, it is singular at
the origin.  Since $g^2\partial_\ta \tau = rF_{tr}$, this singular
soliton is nothing but the electric field outside of a point electric
charge of charge $q$.  Unless electrically charged fields are actually
added to the theory, this solution plays no dynamical role.  In three
dimensions, the energy in the isotropic electric field around a point charge is
logarithmically divergent; strictly speaking we should consider only
configurations with total charge zero.  This is the logarithmic
confinement of ${\rm QED}_3$.

\subsection{The linearly confining phase}

 We may make things more interesting by modifying the theory through
the terms $\Delta\Lgr = \int d^2\ta\ W(T) + c.c.$ with $W=h(e^{pT/2} +
e^{-pT/2})$, making it a sine-Gordon model.  If the radius of the
$U(1)$ gauge group is $2\pi$, then $p$ must be an integer.  For a
trivial K\"ahler potential, this generates a scalar potential $V(T) =
|p\sinh{pT\over 2}|^2$, which has $p$ supersymmetric vacua, located at
$\vev{T}={2\pi i n/p}$, $n$ an integer.  (Note that the positions of
these vacua do not change for a non-trivial but non-singular K\"ahler
potential; the potential $V(T)$ will remain periodic in $T\rarr T+2\pi
i/p$.)  This implies the existence of domain walls separating regions
in different vacua.  Consider a configuration with vacuum $n_1$ at
$x\ll 0$ and vacuum $n_2$ at $x\gg 0$.  There is a non-singular
solution $T(x,y,t)=T(x)$ which interpolates between these two vacua,
giving a domain wall --- a string in two spatial dimensions --- in the
$y$ plane.  Since $\partial_x\tau=F_{ty}$, the string carries electric
flux: it is a linearly confining electric flux tube.  The total flux
in the tube is $\int\ dx F_{ty} = \Delta \tau = {2\pi \over p}$,
and the width of the tube is proportional to $1/p\sqrt{h}.$

Equivalently, note that the existence of $W(T)\neq 0$ changes the
equations so that a constant electric field $F_{xt}(x,y,t) = E$ is no
longer a solution; the equations of motion reduce in this case to
\be{contradiction}
{\partial S(F_{xt})\over \partial F^{xt}}+\partial_y\tau = 0 \ ; \
{df\over d\tau}=0
\ee
the second of which implies $\tau$ is everywhere at a minimum of $f$,
while the first implies $\tau$ must vary linearly with $y$,
inconsistent with the second if $f$ depends on $\tau$.  The electric
field between two parallel line charges will therefore not be uniform,
but will instead break up into strings of electric flux.

\begin{figure}
\centering
\epsfxsize=2.6in
\hspace*{0in}\vspace*{.2in}
\epsffile{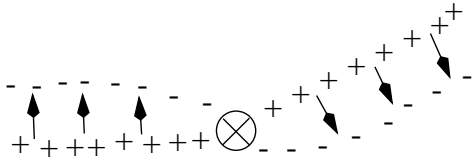}
\caption{The magnetic field around a wire of current is
shielded by a magnetic plasma, and confined into narrow walls.}
\label{fig:plasma}
\end{figure}

Of course, Polyakov explained long ago why the above modification of
the action causes linear confinement.\cite{polymono} The action of
this theory contains the terms ${1\over 2\pi}\int d^3x\ [{1\over
2\pi}\tau dF +f(\tau)].$ Let us work in Euclidean space for a moment.
The Bianchi identity has become $dF=2\pi f'(\tau)$.  Thus, at any
$\tau$ which is not a local minimum or maximum of $f(\tau)$, there is
a local density of magnetic monopoles in the vacuum.  This means that
the vacuum is a magnetic plasma.  An attempt to force a nontrivial
electric field $F_{xt}$ through the Minkowski vacuum by introducing an
electric source corresponds in the Euclidean description to trying to
force a magnetic field through a magnetic plasma by introducing a wire
carrying electric current.  Such a field will of course be screened by
magnetic charge separation (corresponding to $df/d\tau$ taking
positive and negative values in various places) taking place in a
region of order the Debye length $\ell_D$ (inversely proportional to
the monopole charge times the square root of the density, $\ell_D \sim
1/(p\sqrt h)$ in our example.)  This immediately leads to linear
confinement of the electric currents; see \FFig{plasma}. In broken
three-dimensional non-supersymmetric $SU(2)$ Yang-Mills
theory~\cite{polymono} the monopole plasma is generated dynamically,
but the low-energy effective $U(1)$ gauge theory is similar to the one
considered here.

\begin{figure}
\centering
\epsfxsize=2.2in
\hspace*{0in}\vspace*{.2in}
\epsffile{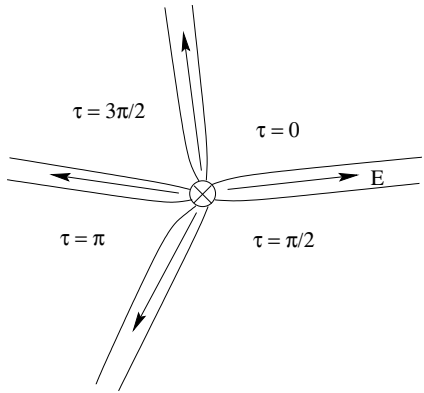}
\caption{The electric field $E$ around an electric charge, represented
by arrows, is confined into walls where $\tau$ varies; between the
walls $\tau$ is found in one of its minima.  This figure applies for
$p=4$.}
\label{fig:hopping}
\end{figure}

Polyakov showed~\cite{polymono} that Wilson loops acquire an area law
when $f(\tau)$ is non-zero and periodic; this is done by considering
the area-dependent effects far from the edge of the loop.\footnote{The
presentation contains many typographical errors; note particularly
that Eq.~(5.23) is the correct solution but neither
(5.22) nor (5.24) is correct as written.}  By considering the edge
effects near the boundary of the loop, the analysis can
straightforwardly be extended to show that for $f(\tau)=0$ Wilson loops
show logarithmic behavior when the kinetic term for $F^\mn$ is
$F^2/g^2$. Similar computations can be
used to show that the Wilson loop has the required behavior in the
conformal phase of SQED with $N_f$ matter fields, where the effective
action for $F$ is $F{\Box}^{-1/2}F$ plus corrections which are small
in the large $N_f$ limit.

\begin{figure}
\centering
\epsfxsize=2.2in
\hspace*{0in}\vspace*{.2in}
\epsffile{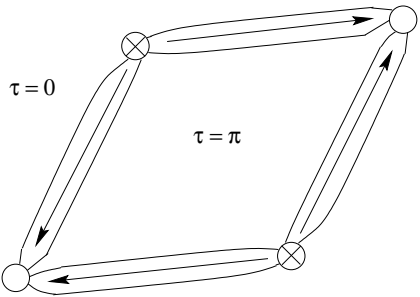}
\caption{A configuration with two heavy particles of
charge 1 and two of charge -1; arrows represent the
confined electric field.}
\label{fig:metastable}
\end{figure}

In this modified theory, the presence of a static point charge cannot
lead to a configuration with $\vev{T(r,\theta,t)}\propto {i\ta}$,
since $\tau$ has $p$ preferred values.  Instead, the point charge,
around which $\tau$ must wind once, will distribute that winding as in
\FFig{hopping}, by hopping from one vacuum to another until $\tau$ has
shifted by $2\pi$.  The electric field $\partial_\ta\tau$ of the
source, instead of being isotropic, is now confined in the flux-carrying
domain walls.  Amusing configurations such as that in
\FFig{metastable} can be constructed; note that similar configurations 
are lacking in QCD only because of the presence of light quarks.

A natural question is whether these domain walls are BPS saturated.
First, consider the question of whether non-trivial BPS bounds exist.
In analogy to kink solitons in two dimensions~\cite{witnolv} and
domain-wall solitons in four dimensions,\cite{dvalshif} BPS bounds for
domain-wall (stringlike) solitons in three dimensions can occur when
the expectation value of the superpotential is different in the two
vacua which the domain wall separates.  Interestingly, because the
superpotential chosen above depends on the vacuum of choice only up to
a sign $W(T=i{ n\pi\over p}) =(-)^n2h$, walls separating vacua with
$n_1-n_2$ odd have a BPS bound while those with $n_1-n_2$ even have no
BPS bound, as in the two-dimensional version of the same
model.\cite{witnolv} This observation suggests that multiple strings
with $n_1-n_2=1$ will bind together to make strings carrying larger
amounts of flux (whose energy cannot be calculated.)  The physics
behind this behavior is not obvious; in particular it is not clear why
the strings in this confining abelian theory are so different from the
domain walls in supersymmetric QCD in four
dimensions,\cite{dvalshif,kss,ewMQCD} which have a BPS bound between
any two vacua.  This also prevents the junction of
multiple strings from being BPS saturated, in contrast to the case of
QCD domain wall junctions \cite{Gibbons,Carroll}.

Bounds having been established for $n_1-n_2$ odd, the next question is
whether such solitons exist.  The BPS-saturated kink solitons for the
sine-Gordon model in two-dimensions have been
constructed,\cite{BPSkink} and can be directly lifted to this
three-dimensional model.

\section{The Abelian Theory with Matter}

\subsection{\ntwo\ Supersymmetric QED}

Consider now \ntwo\ supersymmetric QED, consisting of a single $U(1)$
vector multiplet $V$ and chiral multiplets $Q,\tilde Q$ of charge
$1,-1$.  The vector multiplet in three dimensions contains a real
scalar $\phi$ in addition to the photon and photino.  The Lagrangian
of the theory is
\be{SQEDlgr}
\Lgr_{SQED} = \int d^4\ta\ 
{1\over 4g^2}\Sigma^2 -\left(Q^\dag e^{2V} Q - \Qt^\dag e^{-2V}\Qt\right) 
- \left[\int d^2\ta\ W( Q,\Qt) + c.c.\right]
\ee
Initially we take the superpotential of the theory $W$
to be zero.

\begin{figure}
\centering
\epsfxsize=2.2in
\hspace*{0in}\vspace*{.2in}
\epsffile{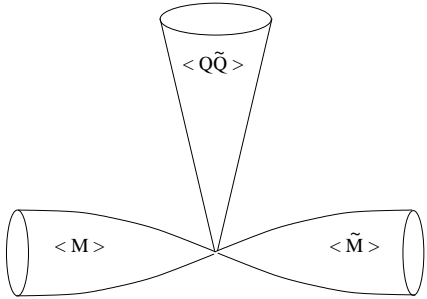}
\caption{The moduli space for \ntwo\ SQED.}
\label{fig:SQEDmodspace}
\end{figure}

The gauge invariant operators of the theory include $Q\tilde Q$, but
the theory also has the gauge-invariant operators $M$ and $\tilde M$
which are sources of magnetic flux.\cite{ntwovort,FT}  The theory
with matter has an abelian Higgs phase $\vev{Q\tilde Q}\neq 0$, in
which the gauge field is massive and the flux which emanates from
$M$ and $\tilde M$ is trapped in magnetic flux vortices, solitons which
behave like particles in three dimensions.  We therefore identify $M$
and $\tilde M$ as vortex-creation operators.\cite{ntwovort,FT}

 The moduli space of the theory consists of the Higgs branch, where
$\vev{Q\tilde Q}\neq 0$ and the gauge group is broken, and the Coulomb
branch, which classically consists of expectation values for the real
field $\phi$.  Quantum mechanically, the Coulomb branch corresponds to
expectation values for $\phi$ and $\tau$, or more precisely for $T$.
For zero superpotential, the Coulomb branch splits into two
pieces,\cite{ntwovort} one on which $M$ has an expectation value, the
other on which $\tilde M$ is nonzero.  At the origin these three
branches touch and there is a conformal field
theory.\cite{ntwovort,ntwobrane} The moduli space is shown in
\FFig{SQEDmodspace}.

\subsection{The XYZ model}

Consider the theory of three chiral superfields $X,Y,Z$ and
superpotential $W=yXYZ$.  The superpotential has dimension 2, while
the fields and $y$ have engineering dimension ${1\over 2}$.  There is
a $U(1)$ R-symmetry under which $W$ has charge $2$ and the fields have
charge $2/3$.

\begin{figure}
\centering
\epsfxsize=2.2in
\hspace*{0in}\vspace*{.2in}
\epsffile{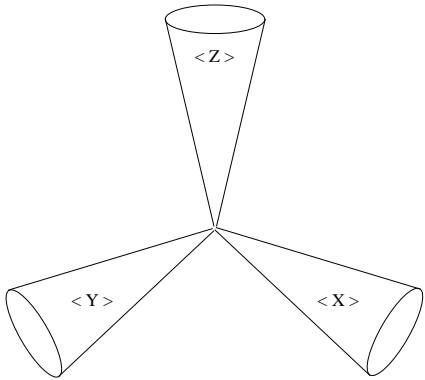}
\caption{The moduli space of the XYZ model.}
\label{fig:threecones}
\end{figure}

The classical scalar potential for the scalar fields,
$V=|(dW/dX)|^2+|(dW/dY)|^2+|(dW/dZ)|^2$, takes the form
$|YZ|^2+|ZX|^2+|XY|^2$.  It has solutions $X\neq 0,Y=Z=0$ and
permutations thereof.  Classically, then, the theory has a space of
vacua consisting of three complex planes, each with one of the three
fields as its coordinate, which touch at the point $X=Y=Z=0$.  At this
special point the R-symmetry and permutation symmetry are unbroken,
and the only scale in the problem is $y$.  It is believed that the
theory flows to a conformal field theory in the infrared.  The
R-charges and permutation symmetry fix the dimensions of $X,Y,Z$ to be
$2/3$.  One way to understand this intuitively is to note that, in the
superpotential, if $y$ were transmuted by quantum dynamics to become a
dimensionless coupling, then $XYZ$ would have dimension 2.  The moduli
space is shown in \FFig{threecones}.

The XYZ model and SQED are surprisingly in the same universality
class.\cite{ntwovort}  For this to be true the theories must have the
same spectrum of gauge invariant operators and the same moduli space,
and indeed they do: the chiral operators $X,Y,Z$ map to the chiral
operators $M,\tilde M,Q\tilde Q$.  The permutation symmetry which is a
classical feature of the XYZ model is a quantum effect of the SQED
conformal field theory.

\subsection{The exact dual of \ntwo\ SQED}
\label{ssec:exactdual}

An exact dual for \nfour\ SQED, valid at all energies, has been
constructed,\cite{FT} and this dual can easily be extended to \ntwo\
SQED.  In particular, it involves coupling a pair of photon multiplets
$V_1,V_2$ with an off-diagonal Chern-Simons term $\int\ d^4\theta\
V_1\Sigma_2$ (also known as a BF term) and using $V_1$ to gauge the
symmetry which rotates the phases of $X$ and $Y$; thus the kinetic
terms become $\int\ d^4\theta\ (X^\dagger e^{V_1} X + Y^\dagger
e^{-V_1} Y)$.  This description is strongly coupled at high energies
(as expected, since SQED is weakly coupled there) and becomes the XYZ
model at very low energies, plus irrelevant operators.  The leading
irrelevant operator is $(X^\dagger X - Y^\dagger Y)^2$.  The addition
of this and other non-singular operators to the K\"ahler potential
does not change the phase of the theory, not does it change the
topological issues which determine the existence of soliton solutions.
For this reason we may use the XYZ model, whose low-energy K\"ahler
potential is unknown in any case, for the investigations below, rather
than the more complicated exact dual description.

The one subtlety which should be noted is that $X^p$ is a gauge
invariant operator in the XYZ model, while in the exact dual of SQED
it is not.  Strictly speaking the operator $X$ in the latter must be
supplemented with a Wilson line $C(\gamma)$, where $\gamma$ is a curve
which extends from the insertion of $X$ to a point at infinity.  Let
us replace $V_2$ with the constrained field $\Sigma_2$ and the
Lagrange multiplier $T_2$ which couples to it by
$(T_2+T_2^\dagger)\Sigma_2$.  In the presence of the $V_1\Sigma_2$
interaction, the field $T_2$ now shifts under the gauge invariance of
$V_1$, and the Wilson line $C(\gamma)$ for $V_1$ along $\gamma$ can be
replaced by $e^{T_2}$.  Consequently the operator $[e^{T_2} X]$, and
any half-integer power of it, is local and gauge invariant and can be
added to the superpotential.  This operator reduces to $X$ when
the gauge fields are integrated out.

\section{The conformal phase}

Let us consider some perturbations of these theories.  The first few
of these are well-studied and serve as tests of the conjectured
infrared equivalence of these theories.  The later ones have not
previously been explored in the literature.  We begin for completeness
with the conformal phase, found at the origin of moduli space.  This
section lies somewhat outside the main flow of the paper and can be
skipped.  The following sections address the phases of central
interest.

\ 

\underline{$W^{XYZ}=yXYZ+t(X^3+Y^3+Z^3)$, 
$W^{SQED}=t(M^3+\tilde M^3+Q^3\tilde Q^3)$}

\ 

Using well known techniques~\cite{emop} it is straightforward to show
that this perturbation is exactly marginal at the origin of moduli
space.\cite{dddmarg}  Specifically, within the space of physical
couplings $h,t$ (which must be distinguished from the holomorphic
couplings appearing in the superpotential above, which do not run)
there is a one-complex-dimensional subspace where the theory remains
conformally invariant.  From the SQED point of view, the
naively-irrelevant tenth-order scalar potential for the charged matter
balances the appearance of a plasma of charge-3 magnetic instantons to
maintain conformal invariance.  Whether this behavior has a simple
physical interpretation is not known.

\

\underline{$W^{XYZ}=yXYZ+tZ^3$, $W^{SQED}=tQ^3\tilde Q^3$}

\

This marginal perturbation is marginally irrelevant. To see this,
consider the beta functions for the theory $W=yXYZ +tZ^3$.
\be{betahy}
\beta_y = y[-\half+\gamma_X(h,t)+\half \gamma_Z(y,t)] \ ; \
\beta_t = t[-\half+{3\over 2}\gamma_Z(y,t)]
\ee
I have used the symmetry between $X,Y$ here to write
$\gamma_Y=\gamma_X$.  Note the following facts.  First, there exists
some $y\equiv y^*$ for which $W=yXYZ$ is conformal; in short,
$y=y^*,t=0$, $\gamma_X=\gamma_Z=\gamma^*=1/3$.  Second,
$\gamma_X(y,t)$, $\gamma_Z(y,t)$ are functions of two variables, and
the conditions in \Eref{betahy} represent two linearly independent
conditions on two variables, whose solutions will be isolated.  The
fixed point $y=y^*,t=0$ is therefore isolated, and thus the
perturbation is not exactly marginal.  To see that it is irrelevant,
note that the solutions to the condition $\gamma_X=\gamma_Z$ form one
or more lines in coupling space. One of these lines, by symmetry, is
at $t=0$ for any value of $y$.  Any other lines will generally lie at
some distance from this one.  For $y=0,t\neq 0$, $\gamma_Z>0$ by
unitarity while $X$ is noninteracting, so $\gamma_Z-\gamma_X$ is
positive.  It must therefore be that for {\it all} $y$ and small $t$,
$\gamma_Z-\gamma_X>0$.  In conclusion, for $y=y^*,t=\epsilon$,
$\beta_t>0$, and so the perturbation is irrelevant.

\

 \underline{$W^{XYZ}=yXYZ+mZ^2$, $W^{SQED}=mQ\tilde 
QQ\tilde Q$}

\

  Integrating out $Z$ leaves a low-energy effective superpotential
$W^{XYZ} = -{y^2\over 4m}XYXY$. This interaction, which gives a sixth-order
scalar potential, is marginally irrelevant in three dimensions.
Consequently, the above perturbation drives the XYZ theory to a free
theory of $X$ and $Y$.  The same perturbation pushes the \ntwo\ SQED
theory to \nfour\ SQED,\cite{dddmarg} which is known to flow to a CFT
that can be written as a free theory of its vortex solitons.

\section{The Logarithmically Confining Phase: non-BPS solitons}

\

\underline{$W^{XYZ}=XYZ+mZ$, $W^{SQED}=mQ\tilde Q$}

\

\begin{figure}
\centering
\epsfxsize=3.2in
\hspace*{0in}\vspace*{.2in}
\epsffile{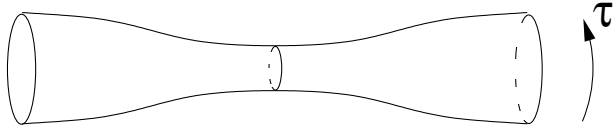}
\caption{The moduli space of SQED with massive matter.  The field
around a massive charged field involves the winding of $\tau$
around the small circle at the center of the space.}
\label{fig:massiveQ}
\end{figure}

A mass for $Q$ and $\tilde Q$ must leave the theory with a moduli
space topologically equivalent to that of a free $U(1)$ gauge theory,
which has only a Coulomb branch with the topology of a
cylinder.\footnote{Henceforth we set $y=1$ for simplicity, since it
plays no role in the following discussion.}  Although the metric
cannot be computed exactly, as would be the case in \nfour\ SQED, it
can be computed using perturbation theory for large $|\phi|$, while
for small $\phi$ it can be constrained using perturbation theory and
our knowledge of the theory with $m=0$.  The radius of the cylinder
asymptotically approaches the gauge coupling. The charged matter
reduces the effective value of the gauge coupling at small $|\phi|$,
and thus the cylinder has smaller radius, as shown in \FFig{massiveQ};
note its consistency with \FFig{SQEDmodspace}.

Because there are no longer any light charged fields in the theory to
screen electric charge, the particles $Q$ and $\tilde Q$ should be
visible as localized massive objects with long range fields.  Actually
this is not quite true; their electric charge means that they are
still logarithmically confined, so they cannot be seen separately from
one another.\footnote{Of course we could avoid the problem of the
logarithmic divergences by weakly breaking the gauge symmetry of the
$U(1)$ theory; this method of regulating the theory will be discussed
below.}  However, if we consider a pair of heavy particles of opposite
charge, separated by a distance $L$, the energy associated with their
electric fields may be much less than the energy associated with their
cores; and so there are states in the theory with energy of order $2
m+\log L\sim 2m$ which have their energy concentrated in two lumps of
size $m^{-1}$ separated by distance $L$.  Can such states be found in
the XYZ theory?

\begin{figure}
\centering
\epsfxsize=3.2in
\hspace*{0in}\vspace*{.2in}
\epsffile{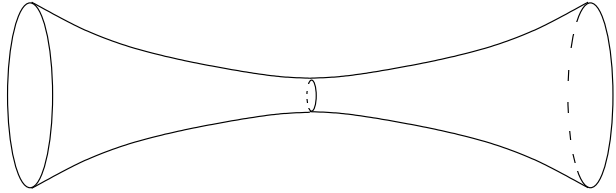}
\caption{The moduli space of the XYZ model with superpotential $W=XYZ+mZ$;
vortex solitons wind around its center.}
\label{fig:hyperbola}
\end{figure}

The answer is yes, and they take the form of logarithmically confined
states of BPS-unsaturated solitons.  The moduli space of the theory
is $XY=m$, a hyperbola, shown in \FFig{hyperbola}.  Since $\pi_1$
of the moduli space is non-zero, there are vortex solitons in which
the fields $X,Y$ wind opposite to one another at spatial infinity.
The kinetic terms for $X,Y$ ensure the energy of one of these solitons
is logarithmically divergent, but a soliton-antisoliton pair would
have finite energy of order twice a core energy plus a long-distance
logarithm.  More importantly, and in contrast to the pure $U(1)$
theory discussed earlier, the core energy is finite; the XYZ model has
solitons because the point at the center of the soliton can have
$X=0,Y=0$ while maintaining finite energy density.
Note that the winding around the hyperbola will be energetically
favored to occur for $|X|=|Y|=\sqrt{|m|}$, the region of maximum
symmetry, since the spatial variation of the fields at infinity, and
the associated stress energy, can be minimized there.

To confirm that the above vortex solitons are the fields $Q$ and
$\tilde Q$, we should check that they are charged under the same
fields.  The winding of the phase of $X$ which defines the soliton
corresponds in SQED to the winding of the phase of $M \sim e^{T}$, and
thus the dual photon $\tau$ shifts by $2\pi$ while winding around the
soliton. Since $\partial_\theta\tau$ is related by electric-magnetic
duality to $F_{tr}$, the radial component of the electric field, the
soliton emits the SQED electric field with total charge $1$.  This
indeed corresponds to the properties of the field $Q$.  The soliton
winding the other way is either $\tilde Q$ or $Q^\dagger$; to
distinguish them one must examine the fermion zero modes of the
solitons more carefully, which although interesting lies outside the
scope of the present paper.

  It is straightforward to verify that these solitons are meaningful,
despite their logarithmic divergences, because the infrared logarithm
may be easily regulated.\footnote{The remainder of this section may be
skipped; it is presented only for technical completeness.}  There are
several options. The easiest is to consider coupling the vector field
$V$ of SQED to a second vector field $\hat V$, using an off-diagonal
Chern-Simons (BF) term $\int \ d^4\ta\ (k/2\pi)\hat V\Sigma$ and
adding a kinetic term $\int \ d^4\ta\ (1/4 g^2)\hat \Sigma^2$.  This
gives a mass $kg^2$ to the vector fields, and means that the
logarithmic potential confining $Q$ and $\tilde Q$ now falls off
exponentially.  The state created by $Q$ alone now has finite energy.
In the $k\rarr\infty$ limit the vector fields decouple, the theory is
free and $Q$ has mass $m$.  In fact, in this limit the theory has
\nfour\ \susy\, and the particle $Q$ is BPS saturated. For small $k$
its mass diverges as $|\log k|$.  The state created by $Q(x)\tilde
Q(0)$ has mass of order $|\log k|$ or $\log g^2|x|$, whichever is
smaller.

The mirror description of this same process requires the exact duality
for SQED described in \SSEC{exactdual}. The finite coupling $g$ in the
original theory corresponds~\cite{FT} to coupling the XYZ model to
topologically massive vector bosons $V_1,V_2$ with mass of order
$g^2$.  The field $\hat V$ couples to $V_2$ through a BF term with
coupling $k$, and it therefore mixes with $V_1$, which couples to
$V_2$ through a BF term with coupling $1$.  The mass matrix for the
vectors $V_1,\hat V$ has a vanishing eigenvalue, so one linear
combination is massive and the other, call it $V_0$, is massless.  The
fields $X,Y$ couple to the massless (massive) vector multiplet with a
coupling proportional to $k$ ($1$), and the massless vector eliminates
the logarithmic divergence in the soliton energy through the
replacement of $\partial_\ta X$ with $(\partial_\ta+A_{0\ta}) X$.  For
$k=\infty$ this modified XYZ theory is simply \nfour\ SQED with a
Fayet-Iliopolous term, which is well-known to be free with massive BPS
vortices.  For finite $k$ the solitons are not BPS saturated but still
have finite mass; only for $k=0$, when the massless vector multiplet
$V_0$ completely decouples from $X,Y$, is the soliton mass divergent.
Meanwhile, for small $k$ and large separation $|x|$, the mass of the
soliton-antisoliton state is of order $\log g^2|x|$, where $g^2$ is
the mass of the topologically massive vector multiplet; it is finite
in the $k\rarr 0$ limit.  This matches with the SQED description
above.

\section{The confining phases:  BPS and non-BPS strings}

\

\underline{$W^{XYZ}=XYZ+mZ+h(X+Y)$, 
$W^{SQED}=mQ\tilde Q+h(M+\tilde M)$}
 
\
 
The presence of linear terms in $X$ and $Y$ implies that we will now
have multiple isolated vacua, each with a mass gap.  This means there
will be regions in different vacua separated by domain walls, as we
discussed earlier for pure $U(1)$ gauge theory.  The vacuum equations
are
\be{XXYYvac}
XY+m=0 \ ; YZ+h=0 \ ; XZ+h=0
\ee
which have two solutions
\be{XXYYsol} 
X=Y=\pm i \sqrt{m}\ ;\ Z = \pm i{h\over \sqrt{m}} 
\ee
These two isolated vacua are two points on the circle
$|X|=|Y|=|\sqrt{m}|$.  If $h$ were zero, then, as described above, a
vortex soliton with logarithmically divergent energy could have been
constructed with asympotic behavior $X(\theta)=i\sqrt{m}e^{i\theta}$,
$Y(\theta)=i\sqrt{m}e^{-i\theta}$.  With $h\neq 0$, only two points of
the moduli space remain, so it is no longer possible to have a soliton
of this type.  But we may ask, what if we had such a soliton for
$h=0$, and then adiabatically turned on a non-zero value for $h$?  The
winding of the soliton cannot be unwound without large energy cost.
Instead, the system will minimize the energy by ensuring that the
phases of $X,Y$ do all of their winding in two narrow domain walls,
sitting in one or the other zero-energy vacuum in the regions between
the walls, as in \FFig{hopping}.  These domain walls, in isolation
(without solitons at the end) are BPS saturated.

\begin{figure}
\centering
\epsfxsize=3.2in
\hspace*{0in}\vspace*{.2in}
\epsffile{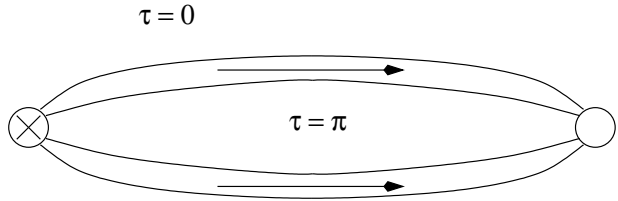}
\caption{In the confining phase with $p=2$, particles with charge 1
and -1 are confined by two strings carrying the minimal quantum of
flux (which most likely bind into a single one with twice the flux.)}
\label{fig:linearconf}
\end{figure}

In short, the vortex soliton will retain its core, but will find
itself at the meeting point of two solitonic flux-carrying domain
walls.  A soliton-antisoliton pair will now be connected by two
strings.  This is shown in \FFig{linearconf}. However, although we
normally expect a pair of identical BPS-saturated solitons to have
zero potential energy, a pair of domain wall solitons of this type is
not BPS saturated, because in fact they are not the same: they connect
different vacua.  Consequently, there is no BPS bound for the pair of
walls shown in \FFig{linearconf} (even in the limit that their endpoints
are taken to infinity), and it is likely the pair of strings in the
figure attract one another and bind to form a single non-BPS string
soliton.

\

\underline{$W^{XYZ}=XYZ+mZ+h(X^{p/2}+Y^{p/2})$,} 

\underline{$W^{SQED}=mQ\tilde Q+h(M^{p/2}+\tilde M^{p/2})$}

\

Here we generalize the previous discussion. The vacuum equations are
\be{XXYYvacpq}
XY+m=0 \ ; YZ+hX^{(p/2)-1}=0 \ ; XZ+hY^{(p/2)-1}=0
\ee
which has $p$ solutions
\be{XXYYsolpq}
X= i \sqrt{m}e^{i\pi n/p}\ ;Y= i \sqrt{m}e^{-i\pi n/p} \ ; \ 
Z =  (-1)^n (i\sqrt{m})^{(p/2)-2}h \ ,  
\ee
where  $n=1,\dots,2p$.
There are thus $p$ isolated vacua on the circle $|X|=|Y|=|\sqrt{m}|$.
In analogy to the case just discussed, this implies that a vortex of
winding number 1 is the meeting place of $p$ flux-carrying domain
walls.

The physics of this case is straightforward. In \eref{XXYYvacpq},
monopoles of charge $p/2$ and density $h$ have been introduced.  Since
the Debye length in the plasma is inversely proportional to $p$,
dimensional analysis shows that the width of a solitonic tube and the
amount of flux it can carry decrease linearly with $p$. To confine an
object of integer electric charge then  requires $p$ such strings.

\

\underline{$W^{XYZ}=XYZ+mZ+hX^{p/2}+\tilde hY^{p/2}$,} 

\underline{$W^{SQED}=mQ\tilde Q+hM^{p/2}+\tilde h\tilde M^{p/2}$}

\

The explicit breaking of the symmetry between $X$ and $Y$
moves the $2p$ vacua. It is easiest to solve for the vacua
by converting the above superpotential back to the
previous one by rescaling $X\rarr AX,Y\rarr Y/A$
where $A=(\tilde h/h)^{1/p}$.  The resulting solutions are
$$
X= i \left({\tilde h\over h}\right)^{1/p}\sqrt{m}e^{i\pi n/p}\ ;
Y= i \left({h\over\tilde  h}\right)^{1/p}\sqrt{m}e^{-i\pi n/p} \ ; \ 
$$
$$
Z =  (-1)^n (i\sqrt{m})^{(p/2)-2}\sqrt{h\tilde h} \ ,
$$
where  $n=1,\dots,2p$.
Notice that in the limit $\tilde h\rarr 0$, $h\neq 0$ the vacua move
to $|Y|=\infty$, $X=0$.  

This limit, for $p=2$, resembles the result of Affleck, Harvey and
Witten~\cite{AHW} for \ntwo\ $SU(2)$ Yang-Mills theory.  In turn this
suggests why linear confinement does not occur in that case.  In
linearly confining SQED we have monopoles $M$ and $\tilde M$ of charge
$1,-1$ and antimonopoles $M^\dagger$ and $\tilde M^\dagger$ of charge
$-1,1$.  If they are both introduced with equal weight as in
\Eref{XXYYvacpq}, then the theory has the symmetry $T\rarr -T$,
implying vacua will be symmetric around $\phi=0$.  Reducing the effect
of charge $-1$ monopoles $\tilde M$ relative to those with charge $1$,
which can be done using the field rescaling given above, shifts
the symmetry to $T\rarr -T+\log (\tilde h/h)^{2/p}$, essentially
acting as a shift of $\phi$. In the limit of infinite rescaling,  
stable supersymmetric vacua have no
natural location other than $\phi=\infty$.  Since in \ntwo\ $SU(2)$
Yang-Mills we find only charge $1$ 't Hooft-Polyakov monopoles $M$ and
charge $-1$ antimonopoles $M^\dagger$, the theory only has a vacuum
for infinite $\vev{\phi}$.

\section{A comment on string theory}

I conclude with some limited remarks concerning the lifting of this
physics to string theory.

  Strings are sources for an abelian two-form gauge field $B^\mn$, and
it is an interesting question whether string theory permits its
electric flux to be confined.  In the context of the present paper,
this is most easily discussed in four dimensions, where the two form
is dual to a scalar $\tau$ (the dimensional reduction of the six-form
which couples to Neveu-Schwarz five-branes.)  In analogy to the
discussion above, one needs to find physics which can generate a
(stable) potential which depends non-trivially on $\tau$.  The
presence of such a potential will indicate that the vacuum has become
a plasma of instantons with magnetic charge under $B^\mn$.  For
example, in four dimensions such instantons would be given by
Euclidean Neveu-Schwarz five-branes wrapped on the six compact
dimensions.

It is easy to construct a four-dimensional field theory model with
such behavior.  The theory of a free string is given by considering
the solitons of the abelian Higgs model, with gauge field $A^\mu$ and
complex Higgs field $\Phi$ and a potential
$V(\Phi)=(\Phi^*\Phi-v^2)^2$. The phase of $\Phi$ --- let us call it
$a$ --- winds around the string.\footnote{Free BPS strings can be
defined in a suitable limit \cite{HSZ} of the pure \ntwo\
supersymmetric Yang-Mills theory studied by Seiberg and
Witten.\cite{nsewone} In the theory with dynamical scale $\Lambda$,
the presence of an \ntwo\ breaking parameter $\mu$ leads to
confinement of electric flux into tubes.  These tubes become \ntwo\
BPS saturated strings in the limit $\mu\rarr 0$, $\Lambda\rarr\infty$
with $\mu\Lambda$ fixed; the parameter $\mu\Lambda$ becomes an
\ntwo-preserving Fayet-Iliopoulos parameter.  The tension of the
strings is $\mu\Lambda$.  Note that, in this context, the phase of the
theory with ``tensionless strings'' consists merely of a photon coupled to
massless uncondensed monopoles at infinite $\Lambda$ --- a free field
theory.}  In fact the low-energy theory well below the scale $v$ can
be written as the gauge field $A^\mu$ coupled to $a$ with the
Lagrangian $(\partial_\mu a + A_\mu)^2$.  Shifting this phase as $a\rarr
a+\Lambda(x)$ is a gauge symmetry.

We may then add a dynamical two-form field $B^\mn$ with a dual scalar
$\tau$, which couples to the magnetic flux in the string soliton via
the interaction $B_\mn F_{\rho\sigma}\epsilon^{\mn\rho\sigma}$.  Since
$\tau$ shifts under the gauge symmetry of $A^\mu$, in analogy to the
discussion in \SSEC{exactdual}, we can define a gauge invariant
operator $\tau+a$, which has a global (not gauged) shift symmetry and
is periodic under shifts by $2\pi$.  The winding of this
gauge-invariant operator around a string soliton corresponds to the
gauge-invariant $B^\mn$ electric flux $H^{0\mn}$ (here $H=dB$), whose
infrared-divergent energy causes logarithmic confinement of strings.
Now, if a non-trivial periodic potential for $\tau+a$ is somehow added
to the theory, then this global shift symmetry is broken.  (For
example, the global shift symmetry of $\tau+a$ can be broken through
an anomaly by instantons in some other gauge group.)  Consequently, as
in \Eref{contradiction}, $H^{0\mn}$ flux is confined, and the string
solitons are linearly confined by axionic domain walls, as in
\FFig{hopping}.

Can a construction along these lines be carried out in a consistent
string theory?  Even if the shift symmetry of $\tau$ is broken, can
the runaway behavior of three-dimensional supersymmetric
Yang-Mills~\cite{AHW} be avoided?  It would be very interesting if a
stable confining phase could be found, especially since string
dualities would lead to a whole class of such phases in which
confinement of various D branes by other (BPS-unsaturated) branes
would occur.

\section*{Acknowledgments}
I thank A. Kapustin, K. Intriligator, V.P. Nair, N. Seiberg and
M. Shifman for conversations. The work reported here was supported in
part by National Science Foundation grant NSF PHY-9513835 and by the
W.M.~Keck Foundation.

\end{document}